\documentclass[a4paper,USenglish]{lipics-v2021}

\usepackage{float}

\newif\ifreviewmode
\newif\ifshowcomments
\newif\ifrevisionmode

\reviewmodefalse
\showcommentsfalse
\revisionmodefalse
\nolinenumbers

\ifreviewmode
  
\else
  
\fi

\ifrevisionmode
  
\else
  
\fi

\ifshowcomments
  \newcommand{\sw}[1]{\textcolor{red}{\textbf{[SW:} #1\textbf{]}}}
\else
  \newcommand{\sw}[1]{}
\fi

\title{Studying Developer Perceptions on the Potential of CI Recommendation Systems}

\author{Osamah H. Alaini}{Department of Computer Science, Trent University, Peterborough, Ontario, Canada} {osamahalaini@trentu.ca}{https://orcid.org/0009-0009-1018-1822}{}

\author{Taher A. Ghaleb}{Department of Computer Science, Trent University, Peterborough, Ontario, Canada} {taherghaleb@trentu.ca}{https://orcid.org/0000-0001-9336-7298}{}

\authorrunning{Osamah H. Alaini and Taher A. Ghaleb}
\Copyright{Osamah H. Alaini and Taher A. Ghaleb}



\usepackage{pifont}
\newcommand{\cmark}{\ding{51}}
\newcommand{\xmark}{\ding{55}}
\newcommand{\deepQ}[2]{
  Q\raisebox{-0.25ex}{\scalebox{0.60}{#1.#2}}
}
\newcommand{\deepA}[2]{
  A\hspace{0.15em}\raisebox{-0.25ex}{\scalebox{0.60}{#1--#2}}
}

\newcommand{\deepAone}[1]{
  A\raisebox{-0.25ex}{\scalebox{0.60}{#1}}
}

\newcommand{\deepBone}[1]{
  B\raisebox{-0.25ex}{\scalebox{0.60}{#1}}
}

\begin{document}

\maketitle

\begin{abstract}
Continuous Integration (CI) is central to modern software development, yet developers often struggle to choose the most suitable CI service. Prior work has identified barriers to CI adoption but offers little empirical evidence on how developers select CI services or whether adoption decisions are driven by genuine project needs versus social influence. This paper presents an exploratory survey study addressing that gap. We aim to contact about 5,000 active GitHub developers, including both CI users and non-users. The study investigates: (1) what drives CI adoption and service selection, distinguishing need-driven from socially influenced motivations; (2) whether developers consider CI universally necessary or context-dependent and what barriers hinder adoption; and (3) developers' perceptions of automated CI recommendation systems. Our findings will inform researchers developing CI recommendation systems and practitioners aiming to streamline CI adoption in open-source projects.
\end{abstract}

\section{Introduction}
\label{sec:introduction}
\vspace{-6pt}
Continuous Integration (CI) is central to modern software development, automating testing and deployment while reducing integration errors~\cite{fowler2006continuous,hilton2017tradeoffs}. CI adoption has increased from about 40\% in open-source projects~\cite{hilton2016usage} to over 50\% with GitHub Actions~\cite{golzadeh2022rise}, yet developers still struggle to choose services amid complex trade-offs in features, pricing, and workflow~\cite{chopra2025multiCI,rostami2023usage}. The impact is substantial: 18.8\% of Java projects abandon CI entirely~\cite{chopra2025multiCI}, 52\% of developers want easier configuration~\cite{hilton2017tradeoffs}, and socially driven adoption rather than project needs wastes time and resources. This challenge is expected to grow with the increasing adoption of AI agents for configuring CI/CD workflows~\cite{ghaleb2026agentcicd}.

Prior work has documented adoption barriers through systematic reviews~\cite{shahin2017systematic}, pain-point taxonomies~\cite{widder2019painpoints}, and analyses of technical and social factors~\cite{pinto2018practices,ghaleb2019duration,ghaleb2022interplay}. Three critical gaps remain: (1) limited understanding of whether adoption decisions are driven by project needs or social influence, (2) no empirical evidence on whether developers view CI as universally necessary or context-dependent and what barriers prevent adoption among those who view it as inappropriate, and (3) no empirical evidence on developers' receptiveness to automated CI recommendation systems that could advise on suitability, suggest compatible services, and assist with configuration based on project context. These gaps are particularly relevant given the desire for easier configuration~\cite{hilton2017tradeoffs} and prior work on CI migration~\cite{hossain2025cigrate}. Together, these gaps motivate the first stage of a broader research agenda toward context-aware, AI-enabled CI support, where project suitability, service selection, and configuration are guided by empirical project characteristics rather than defaults or social influence~\cite{alaini2026vision}.

This paper addresses these gaps by surveying about 5,000 active GitHub developers (expected responses: 250–500). Unlike prior surveys targeting only CI users~\cite{hilton2016usage,pinto2018practices} or experienced adopters~\cite{rostami2023usage}, we explicitly include non-adopters to enable comparative analysis across three RQs:
(RQ1) What drives CI adoption and service selection, distinguishing need-driven from socially-influenced motivations? 
(RQ2) Do developers view CI as universally necessary or context-dependent, and what barriers prevent adoption? 
(RQ3) Do developers perceive value in CI recommendation systems, and what features do they desire? 
Specifically, we examine perceived value across three dimensions: assessing project suitability for CI, recommending services given project characteristics (e.g., language, team size, testing practices), and supporting configuration and setup.

The rest of this paper is organized as follows. Section~\ref{sec:background} reviews related work on adoption barriers, service selection, and CI recommendation systems. Section~\ref{sec:study_design} outlines the study design, sampling strategy, survey instrument, analysis procedures, execution plan, ethical considerations, and threats to validity. Section~\ref{sec:conclusion} concludes the paper.

\section{Background and Related Work}
\label{sec:background}

\subsection{Background}

\subsubsection{Continuous Integration: Overview}
Continuous Integration (CI) is a software development practice in which developers frequently integrate code changes into a shared repository, triggering automated builds and tests to detect integration issues early~\cite{fowler2006continuous}. A typical CI workflow involves: (1) committing code changes to version control, (2) automatically retrieving and building the latest code, (3) executing automated tests, and (4) reporting build status to the team~\cite{fowler2006continuous}.

\subsubsection{CI Services and Ecosystem}
Modern CI platforms fall into two categories: self-hosted solutions (e.g., Jenkins), which require organizations to maintain their own infrastructure, and cloud-based services (e.g., Travis CI, CircleCI, GitHub Actions, AppVeyor), which integrate directly with version control platforms. Before 2019, CI research focused mainly on Travis CI. Since then, the landscape has shifted with GitHub Actions~\cite{decan2022github,golzadeh2022rise}, which integrates tightly with GitHub repositories, provides a marketplace of reusable workflow components, and supports matrix builds across multiple environments~\cite{decan2022github}. This diverse ecosystem forces developers to make complex choices, balancing pricing models, feature sets (e.g., parallelization, caching, deployment integrations), platform support, and configuration complexity~\cite{chopra2025multiCI,rostami2023usage}.

\subsubsection{CI Principles and Practices}
CI emphasizes automated testing, fast feedback loops, and continuous deployment~\cite{fowler2006continuous,humble2010cd}, with workflows typically covering dependency installation, compilation, unit and integration tests, and static analysis. Despite these standardized principles, open-source projects exhibit considerable inconsistencies in CI configurations~\cite{ghaleb2025android,abrokwah2025complexity,abrokwah2026compliant}. Modern CI ecosystems also integrate security tooling; for example, \texttt{codeql} is among the most common workflow names, appearing in 3.5\% of GitHub Actions repositories~\cite{decan2022github}.

\subsection{Related Work}

\subsubsection{CI Adoption Barriers and Service Selection}
Prior research has documented persistent adoption challenges. Hilton et al.~\cite{hilton2016usage} surveyed 442 developers and identified setup difficulties, configuration complexity, and lack of experience as key barriers. Pinto et al.~\cite{pinto2018practices} surveyed 158 Travis CI users and found that 33.1\% were uncertain whether job failures constituted build failures, revealing conceptual gaps. Widder et al.~\cite{widder2019painpoints} replicated these results, confirming that these barriers persist. Overall, these studies show that adoption barriers are not only technical but also knowledge-based and contextual. Beyond adoption, developers may also struggle with complex service selection decisions.
Chopra and Ghaleb~\cite{chopra2025multiCI} analyzed 19K Java projects and found that 64\% rely on a single CI maintainer (creating knowledge bottlenecks) and 18\% use multiple CI services (indicating uncertainty about which service fits their project). Travis CI abandonment (29.3\%) exceeded the overall rate (18.8\%), but they did not study CI recommendation system adoption. 
Gallaba and McIntosh~\cite{gallaba2020misuse} reported that 48.16\% of CI configuration code configures job processing nodes and 9.60\% of projects contain anti-patterns. Prior work has also shown trade-offs between build speed and reliability~\cite{ghaleb2019duration,ghaleb2022interplay,ghaleb2026promise}.
These findings directly motivate our research: understanding what drives adoption and service selection, distinguishing need-driven from socially-influenced motivations (RQ1), whether developers view CI as universally necessary or context-dependent and what barriers prevent adoption (RQ2), and assessing developers' perceptions on developing CI recommendation systems (RQ3).

\subsubsection{Qualitative Studies of CI Migration} 
\vspace{-2pt}
Complementing quantitative surveys, Mazrae et al.~\cite{rostami2023usage} conducted interviews with 22 practitioners working with 31 CI tools, revealing that migration decisions stem from evolving project needs, service discontinuation, or improved platform integration. Despite documenting migration patterns, developer perceptions on developing automated CI recommendation systems remain unexplored.

\subsubsection{Tool-Development Efforts and Research Gap}
\vspace{-2pt}
Researchers have developed automated solutions to address CI challenges, including tools for CI service migration~\cite{hossain2025cigrate} approaches addressing CI/CD workflow resource usage and maintenance~\cite{valenzuela2024automation,bouzenia2024resource}. While implementing CI recommendation systems involves technical challenges, such as extracting project characteristics and maintaining service knowledge, these efforts suggest that the underlying components are feasible.
Despite this progress, little is known about whether developers would adopt automated CI recommendation systems or which features they would value. Understanding these factors is critical, as technology acceptance models (TAM~\cite{davis1989tam} and TAM2~\cite{venkatesh2000tam2}) suggest that perceived usefulness strongly influence whether practitioners adopt new tools, while trust is also widely recognized as an important factor shaping technology adoption decisions. Without empirical evidence on these aspects, future tool development risks becoming misaligned with developer needs. 
Hence, we ground RQ3 in the TAM constructs of perceived usefulness and behavioral intention, together with trust to evaluate developers' anticipated acceptance of CI recommendation systems.

\subsubsection{Prior Developer Surveys}
Several large-scale CI surveys have been conducted.
Table~\ref{tab:survey_comparison} compares our study with prior CI surveys across key research dimensions.
Hilton et al.~\cite{hilton2016usage} surveyed 442 developers; Pinto et al.~\cite{pinto2018practices} surveyed 158 Travis CI users; Widder et al.~\cite{widder2019painpoints} replicated Hilton's methodology; Elazhary et al.~\cite{elazhary2021benefits} conducted organizational interviews.
While these studies documented barriers and practices, they share three methodological limitations that our study addresses: (1) minimal representation of non-adopters (Hilton et al.\ surveyed only 7.9\% non-users, while Pinto et al., Widder et al., and Elazhary et al.\ focused exclusively on CI users), which we address via inclusive sampling of both CI users and non-users; (2) no assessment of developers' perceptions on CI recommendation systems despite well-documented configuration challenges (e.g., 52\% desire easier configuration~\cite{hilton2017tradeoffs}), which we address via direct Likert-scale measurement of perceived value and desired features; and (3) the absence of a comprehensive survey of the post-GitHub Actions (2019) CI ecosystem, which we address by surveying the current landscape across RQ1--RQ3.

\begin{table}[htbp]
\centering
\setlength{\tabcolsep}{3.8pt}
\renewcommand{\arraystretch}{1}
\captionsetup{font=footnotesize, labelfont=bf}
\caption{Coverage of CI/CD surveys across key dimensions. Earlier work examined barriers but not service selection, need vs. social influence, or recommendation systems}
\label{tab:survey_comparison}
\setlength{\arrayrulewidth}{0.4pt} 

\vspace{-7pt}
\resizebox{\textwidth}{!}{
\begin{tabular}{l|p{1.5cm}|p{1.3cm}|p{1.3cm}|p{1.3cm}|p{1.4cm}|p{1.4cm}|p{1.8cm}}
\hline
\textbf{Question/Objective} & \textbf{Hilton 2016} & \textbf{Hilton 2017} & \textbf{Pinto 2018} & \textbf{Widder 2019} & \textbf{Elazhary 2021} & \textbf{Rostami 2023} & \textbf{This Study} \\
\hline

CI adoption barriers & \cmark (pre-GHA) & \cmark (pre-GHA) & \cmark (pre-GHA) & \cmark (pre-GHA) & \cmark (pre-GHA) & Partial & \cmark (post-GHA) \\
\hline

Service selection criteria & \xmark & \xmark & \xmark & \xmark & \xmark & \cmark (n=22) & \cmark (n=250--500) \\
\hline

Need vs. social influence & \xmark & \xmark & \xmark & \xmark & \xmark & Partial & \cmark (explicit) \\
\hline

Multi-CI usage reasons & \xmark & \xmark & \xmark & \xmark & \xmark & \cmark & \cmark \\
\hline

Service migration/switching & \xmark & \xmark & \xmark & \cmark (leavers) & \xmark & \cmark (n=22) & \cmark (n=250--500) \\
\hline

Non-user perspectives & Limited (7.9\%) & \xmark & \xmark & \xmark & \xmark & \xmark & \cmark (40--60\%) \\
\hline

Developer support for CI recommendation systems & \xmark & \xmark & \xmark & \xmark & \xmark & \xmark & \cmark \\
\hline

Post-GitHub Actions era & \xmark & \xmark & \xmark & \xmark & \xmark & \cmark & \cmark \\
\hline

\end{tabular}
}
\vspace{-5pt}
\end{table}

\section{Study Design and Execution Plan}
\label{sec:study_design}

\subsection{Research Questions and Methodology}
\begin{figure*}[t]
\centering
\vspace{-5pt}
\includegraphics[width=\textwidth]{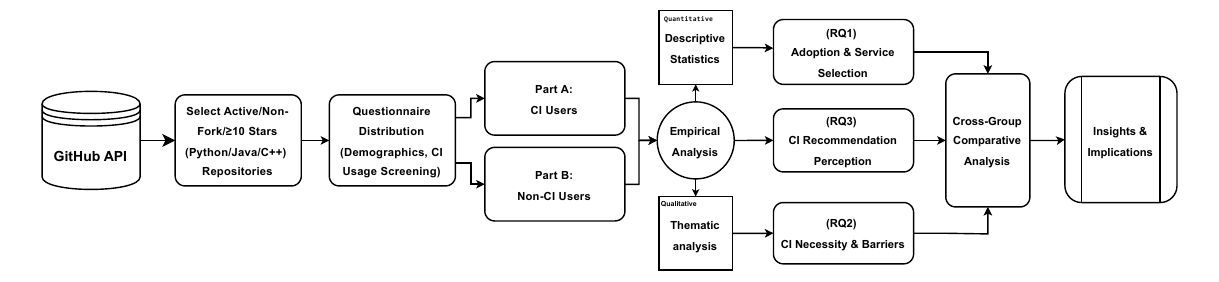}
\vspace{-15pt}
\caption{Overview of our survey study design}
\label{fig:study_overview}
\vspace{-10pt}
\end{figure*}

We will conduct a survey-based empirical study following established practices from prior CI surveys~\cite{hilton2016usage,pinto2018practices} and GitHub mining guidelines~\cite{kalliamvakou2014promises} and general empirical software engineering methodology~\cite{wohlin2012experimentation}. Respondents will be segmented by CI involvement status (CI users versus non-CI users) based on their self-declaration. In this study, CI users are respondents who report having configured, maintained, or actively used CI in a software project, whereas non-CI users report no such involvement. This segmentation enables comparative analyses, which are essential for understanding developer needs in the development of CI recommendation systems. Based on prior evidence that approximately 40\% of GitHub projects use CI~\cite{hilton2016usage}, and more recent studies indicating adoption above 50\% in the post-GitHub Actions era~\cite{golzadeh2022rise}, we anticipate our sample will naturally reflect a distribution within this range.   
Figure~\ref{fig:study_overview} provides an overview of our methodology. In particular, we address three research questions:

\smallskip
\noindent\textbf{RQ1: What factors drive CI adoption and service selection, and how do project needs versus social influence shape these decisions?}

    We investigate three dimensions:
    \begin{itemize}
        \item \textbf{RQ1.1:} What motivates initial CI adoption, and to what extent are they driven by project-specific needs versus social influences (e.g., following popular projects)?
        \item \textbf{RQ1.2:} Which criteria do developers prioritize when choosing specific CI services (e.g., ease of use, cost, features, integration, reliability)?  
        \item \textbf{RQ1.3:} How do developers switch between CI services or adopt multiple services concurrently, and what factors influence these decisions?
    \end{itemize}

Understanding these factors helps identify which adoption motivations and service features a CI recommendation system should prioritize.

\smallskip
\noindent\textbf{RQ2: How do developers perceive the necessity of CI, and what contextual factors limit its adoption?}

We investigate three dimensions:
    \begin{itemize}
        \item \textbf{RQ2.1:} Do developers consider CI essential for all projects, or beneficial only in specific contexts?  
        \item \textbf{RQ2.2:} What reasoning guides developers in choosing to adopt CI selectively rather than universally?  
        \item \textbf{RQ2.3:} What types of barriers prevent CI adoption among non-adopters, and how prevalent is each barrier type?
        \end{itemize}

Understanding whether CI is seen as universally necessary or context-dependent informs whether CI recommendation systems should suggest CI for all projects or first assess each project's suitability.

\smallskip
\noindent\textbf{RQ3: How do developers perceive the usefulness and desired features of a CI recommendation system?}

We aim to measure the perceived value using Likert scales in terms of usefulness, trust, and likelihood of using recommendations. Desired needs are captured via multi-select checkboxes (e.g., service comparison, configuration templates, troubleshooting guides, integration examples) and open-ended responses, enabling identification of actionable requirements for developing such systems.
Based on prior findings that 52\% of developers desire easier configuration \cite{hilton2017tradeoffs} and documented service selection challenges \cite{chopra2025multiCI}, we anticipate exploring patterns in perceived value across different developer groups.

\subsection{Study Design}

\noindent\textbf{Target population and sampling.}
Following GitHub mining best practices~\cite{kalliamvakou2014promises}, we will use the GitHub REST API to sample about 5,000 active open-source developers with publicly available email addresses from repositories that have at least 10 stars and 10 commits in the past 12 months.
For each candidate repository, we will identify recent committing developers from the past 12 months and randomly select one developer. We will exclude bots and accounts with non-contactable emails. To prevent repeated invitations to the same person, we will remove duplicated email addresses across all repositories, if any.
This 5,000 represents our \emph{sampling population}, i.e., the number of invitation emails to be sent, not the expected number of completed responses. We aim to receive 250–500, as detailed below.

\smallskip
\noindent\textbf{Prospective response rate.}
We aim for 250-500 complete responses (5-10\% response rate), which is motivated by prior CI developer surveys: Hilton et al.~\cite{hilton2016usage} achieved 9.8\% (442 of 4,508), Pinto et al.~\cite{pinto2018practices} achieved 14.4\% (158 of 1,100), and Ghaleb et al.~\cite{ghaleb2022interplay} achieved 3.5\% (139 of 4,366). Hence, we aim for the middle of this range (5-10\%) following established strategies: personalized outreach, optimized survey length ($\approx$15 minutes), and mobile-responsive design. Research shows these strategies can collectively increase response rate in developer surveys~\cite{dillman2014internet}.
We acknowledge that non-CI developers may respond at lower rates, so we will monitor response rate throughout the process and may prioritize non-CI developers, if needed, in the subsequent rounds of invitations.

\smallskip
\noindent\textbf{Expected sample composition}
Prior work estimates CI adoption in GitHub projects at 40\%~\cite{hilton2016usage} to over 50\% in the post-GitHub Actions era~\cite{golzadeh2022rise}. We therefore conservatively expect 40--60\% CI users in our sample. For a minimum sample of 250 developers, this yields about 100 CI and 150 non-CI users; for our 500-developer target, about 200 CI and 300 non-CI users. This distribution provides adequate power for subgroup analyses: Mann-Whitney $U$ tests with a medium effect size ($r = 0.3$) require $n \ge 87$ per group for 80\% power at $\alpha = 0.05$~\cite{cohen1988statistical}, which both sample sizes satisfy.

\smallskip
\noindent\textbf{Survey instrument:} We used a two-part questionnaire with conditional routing implemented in Qualtrics.\footnote{\url{https://trentu.qualtrics.com/jfe/form/SV_0fbtOMEDAyNQpam}}
Table~\ref{tab:survey-instrument} gives an overview of the types of questions and answers included at a high-level (full survey instrument is available online\footnote{\url{https://docs.google.com/document/d/1xcDl2n1Q09fgpUvvWLDnVPnimkqcMmYi_UQ71FlTpZs}}).

Participants indicate whether they have been involved with CI (including configuring, maintaining, or actively using CI), routing them to the appropriate part. To ensure consistent interpretation of "CI recommendation systems," the survey presents respondents with the explicit definition provided in Section~\ref{sec:introduction}.

\begin{itemize}
    \vspace{-5pt}
    \item \textit{\textbf{Part A (CI Users, ~18 questions)}} focuses on adoption motivations, service selection, and switching (RQ1.1, RQ1.2, RQ1.3), opinions about CI universality and selective adoption (RQ2.1, RQ2.2), and perceived value of CI recommendation systems (RQ3).
    
    \item \textit{\textbf{Part B (Non-CI Users, ~14 questions)}} explores views on CI universality (RQ2.1), adoption barriers (RQ2.3), and perceived value of CI recommendation systems (RQ3).
    \vspace{-5pt}
\end{itemize}

\smallskip
\noindent\textbf{Perceived value.}
Both \textit{Part A} and \textit{Part B} include 3-4 Likert questions measuring perceived value: (1) usefulness (``Would help me''), (2) trust (``I would trust recommendations''), (3) adoption likelihood (``I would use a CI recommendation system'').  Desired needs are captured via a multi-select list (service comparison, configuration templates, troubleshooting guides, integration examples, performance benchmarks, community feedback) and an open-ended text field for additional features.

\subsection{Quantitative Analysis}
\textbf{RQ1 (adoption motivations):}
We will compute descriptive statistics (frequencies, medians, interquartiles) and compare need-driven versus socially-influenced motivation distributions via Chi-square tests. We will also Mann-Whitney $U$ tests~\cite{mann1947test} (effect size: $r$)~\cite{grissom2005effect} to compare ordinal Likert outcomes. For RQ1.2 (service selection criteria), we will analyze frequencies and crosstabs of prioritized criteria. For RQ1.3 (switching/multi-service adoption), we will compute frequencies and conduct thematic analysis of open-ended explanations.

\begin{table}[H]
\centering
\setlength{\tabcolsep}{3.8pt}
\renewcommand{\arraystretch}{1.06}
\captionsetup{font=footnotesize, labelfont=bf}
\caption{\textcolor{black}{Survey instrument mapping questions to research questions (RQ1: adoption factors; RQ2: barriers; RQ3: perceived value)}}
\vspace{-5pt}
\label{tab:survey-instrument}
\setlength{\arrayrulewidth}{0.01pt}
\resizebox{1.07\textwidth}{!}{\color{black}
    \begin{tabular}{p{0.05\textwidth}|p{0.55\textwidth}|p{0.57\textwidth}|p{0.05\textwidth}}
    
    \hline
    \textbf{\#} & \textbf{Question} & \textbf{Answer Options} & \textbf{RQ} \\
    \hline
    
    \multicolumn{4}{l}{\textbf{Section 1: Demographics (5 items)}}\\
    \hline
    \deepQ{1}{1} & Which country do you work from? 
         & Open-ended (country) & -- \\
    \hline
    \deepQ{1}{2} & What is your age group? 
         & [Under 24, 25--34, 35--44, 45--54, 55+] & -- \\
    \hline
    \deepQ{1}{3} & What is your gender? 
         & [Man, Woman, Non-binary, Prefer not to say] & -- \\
    \hline
    \deepQ{1}{4} & How many years of experience do you have in software development?
         & [$<1$, 1--3, 4--6, 7--10, 10+ years] & -- \\
    \hline
    \deepQ{1}{5} & Which best describes your current role? 
         & [Engineer, Student, Researcher, DevOps, OSS contributor, Other (open-ended)] & -- \\
    \hline
    
    \multicolumn{4}{l}{\textbf{Section 2: CI Usage Patterns (routing item)}}\\
    \hline
    \deepQ{2}{1} & Have you ever been involved with Continuous Integration (CI) in any of your software projects? This includes: configuring or setting up CI services (e.g., writing .yml files, creating workflows); maintaining or modifying existing CI configurations; and actively using CI (triggering builds, reviewing results, fixing failures). 
         & [Yes $\rightarrow$ Part A; No $\rightarrow$ Part B] & -- \\
    \hline
    
    \multicolumn{4}{l}{\textbf{Part A: CI adopters (18 items, A1--A18)}}\\
    \hline
    \deepA{1}{3}
      & CI experience, scope of adoption, and services adopted
      & Multiple-choice (experience bands; all vs.\ some projects; CI services adopted, select all + ``Other'') & -- \\
    \hline
    \deepAone{4}
      & Primary reason for CI adoption
      & Single-choice (improve quality, catch bugs, team requirement, best practice, automate tasks, recommended, deployment, ``Other'') & RQ1 \\
    \hline
    \deepAone{5}
      & CI service consistency across projects
      & Single-choice (same/different) + open-ended justification & RQ1 \\
    \hline
    \deepA{6}{7}
      & Service selection criteria consideration
      & Yes/No + conditional select all that apply & RQ1 \\
    \hline
    \deepAone{8}
      & Perceived usefulness of CI recommendation tool
      & 5-point Likert (1: not useful at all \dots 5: very useful) & RQ3 \\
    \hline
    \deepA{9}{10}
      & CI adoption challenges and specific pain points
      & 5-point Likert (overall difficulty) + select up to 3 challenges (build times, flaky tests, config complexity, debugging, timeouts, cost, documentation, integration, team resistance, learning curve, maintenance, none, ``Other'') & RQ1 \\
    \hline
    \deepA{11}{12}
      & CI service switching and multi-service usage patterns
      & Yes/No + open-ended justification (for each) & RQ1 \\
    \hline
    \deepA{13}{14}
      & Opinions about CI necessity and adoption suggestions
      & Yes/No + open-ended justification; open-ended feedback & RQ2 \\
    \hline
    \deepAone{15}
      & Satisfaction with current CI service selection
      & 5-point Likert (1: very dissatisfied \dots 5: very satisfied) + conditional open-ended (if 1--3: ``What would make you more satisfied?'') & RQ1 \\
    \hline
    \deepAone{16}
      & Trust in automated CI recommendation system
      & 5-point Likert (1: no trust at all \dots 5: complete trust) & RQ3 \\
    \hline
    \deepAone{17}
      & Likelihood to use CI recommendation system
      & 5-point Likert (1: very unlikely \dots 5: very likely) & RQ3 \\
    \hline
    \deepAone{18}
      & Desired capabilities in CI recommendation system
      & Select all that apply  & RQ3 \\
    \hline
    
    \multicolumn{4}{l}{\textbf{Part B: Non-CI adopters (14 items, B1--B14)}}\\
    \hline
    \deepBone{1}
      & Familiarity with the concept of CI 
      & 5-point Likert (1: not familiar at all \dots 5: very familiar) & -- \\
    \hline
    \deepBone{2}
      & Reasons for not adopting CI so far
      & Select all that apply & RQ2 \\
    \hline
    \deepBone{3}
      & Single biggest barrier preventing CI adoption
      & Single-choice (lack of knowledge, time constraints, project complexity, cost, setup complexity, no perceived need, no team requirement, service uncertainty, ``Other'') & RQ2 \\
    \hline
    \deepBone{4}
      & Beliefs about CI necessity for all projects
      & Yes/No + open-ended justification & RQ2 \\
    \hline
    \deepBone{5}
      & Future adoption likelihood if easier setup
      & 5-point Likert (1: very unlikely \dots 5: very likely) & RQ2 \\
    \hline
    \deepBone{6}
      & Factors that would motivate CI adoption
      & Select all that apply & RQ2 \\
    \hline
    \deepBone{7}
      & Current code checking process before deployment
      & Select all that apply & -- \\
    \hline
    \deepBone{8}
      & Receptiveness to CI recommendation system
      & 5-point Likert (1: not interested at all \dots 5: very interested) & RQ3 \\
    \hline
    \deepBone{9}
      & Concerns, doubts, and feedback about CI adoption
      & Open-ended response & RQ2 \\
    \hline
    \deepBone{10}
      & Additional thoughts about CI
      & Open-ended response & -- \\
    \hline
    \deepBone{11}
      & Perceived effort to set up and maintain CI
      & 5-point Likert (1: very little effort \dots 5: very significant effort) & RQ2 \\
    \hline
    \deepBone{12}
      & Trust in automated CI recommendation system
      & 5-point Likert (1: no trust at all \dots 5: complete trust) & RQ3 \\
    \hline
    \deepBone{13}
      & Likelihood to use CI recommendation system
      & 5-point Likert (1: very unlikely \dots 5: very likely) & RQ3 \\
    \hline
    \deepBone{14}
      & Desired capabilities in CI recommendation system
      & Select all that apply & RQ3 \\
    \hline
    \end{tabular}
    }
\end{table}

\noindent\textbf{RQ2 (CI universality patterns):}
We will compare response frequencies on whether developers believe all projects should adopt CI between CI users and non-users using Chi-square tests. For RQ2.2 (reasons for selective adoption), we will thematically analyze open-ended responses from developers who adopt CI selectively. For RQ2.3 (barriers among context-dependent believers), we will compute descriptive statistics for informational, technical, and organizational barriers and compare subgroups using Mann-Whitney U tests.

\smallskip
\noindent\textbf{RQ3 (perceived value):} Descriptive statistics for perceived value Likert questions (mean, SD, frequencies). Mann-Whitney U tests comparing ratings between CI users (Part A) and non-CI users (Part B; effect size: $r$). Frequency analysis of multi-select developer needs.

\subsection{Qualitative Analysis of Open-Ended Responses}
\vspace{-2pt}
We will conduct an in-depth thematic analysis following Braun and Clarke~\cite{braun2006thematic} using a hybrid coding approach. Initial codes will be informed by prior CI research~\cite{hilton2016usage,hilton2017tradeoffs,widder2019painpoints,pinto2018practices}, while remaining open to emergent themes. The two co-authors (software engineering researchers experienced in CI) will collaboratively code all responses to ensure consistency.
We will monitor thematic saturation in sequential batches, declaring it when no new codes appear across consecutive batches. The coders will jointly refine a coding scheme organized by research question (RQ1: adoption motivations, service selection trade-offs, switching triggers; RQ2: universality patterns, selective adoption reasoning, context-dependent barriers; RQ3: trust concerns, desired features). We will assess inter-rater reliability using Cohen’s kappa and resolve discrepancies through discussion.
The final scheme will be applied to the full dataset, with 100\% of coded responses reviewed for quality assurance. We will extract code frequencies and illustrative quotes, segmented by CI user status.

\subsection{Execution Plan}
\vspace{-2pt}
\noindent\textbf{Phase 1: Research Ethics Approval and Setup.}
We have prepared the full survey protocol with all methodological and ethical details and submitted it for ethics approval. After approval, we will update the Qualtrics survey as required and run a pilot with 10–15 developers to estimate completion time and test technical functionality. We will refine the survey based on pilot feedback, so the final wording and number of questions may differ slightly from those proposed here, while preserving the same constructs and RQ mappings.

\smallskip
\noindent\textbf{Phase 2: Developer Sampling.}
We will use Python scripts to query the GitHub API with star and activity filters; validate and de-duplicate email addresses; remove bots and invalid emails; and organize the final $\approx5,000$ invitations by language, geography, and experience level for balanced distribution.

\smallskip
\noindent\textbf{Phase 3: Survey Distribution.}
We will send personalized email invitations to developers in several rounds, track responses, send one generic reminder (to be ignored by those who already responded), and target 250–500 total responses. If response rate is lower than expected, we may add a small random-draw incentive, consistent with prior CI research~\cite{hilton2016usage,ghaleb2022interplay}.

\smallskip
\noindent\textbf{Phase 4: Data Cleaning.}
We will export anonymized responses;
identify incomplete responses (threshold: <80\% questions; document exclusion); validate final dataset; generate demographic profile; assess representativeness.

\smallskip
\noindent\textbf{Phase 5: Quantitative Analysis.}
We will compute descriptive statistics (frequencies, medians, interquartiles) for all Likert questions by CI user status and subgroups (experience, role, project size); perform Mann–Whitney U tests on key questions (e.g., RQ3 usefulness), reporting p-values and effect sizes; create contingency tables for categorical questions and run Chi-square tests; produce visualizations (e.g., box plots and stacked bar charts).

\smallskip
\noindent\textbf{Phase 6: Qualitative Analysis.}
We will develop coding scheme via initial subset coding (10\%) with inter-rater agreement validation (target kappa $\geq$ 0.70); apply the scheme to the full dataset with spot-check (second coder reviews 20\% of remaining responses). Compute code frequencies; extract representative quotes (2-5 per theme); cross-tabulate themes by CI user status and demographic subgroups.

\smallskip
\noindent\textbf{Phase 7: Integration and Synthesis.}
We will triangulate quantitative and qualitative findings for each RQ: RQ1 (adoption motivations, service selection, switching patterns), RQ2 (universality patterns, selective adoption reasoning, barriers among context-dependent believers), and RQ3 (perceived usefulness ratings with qualitative statements on trust and needs). RQ3 Likert items will follow the Technology Acceptance Model~\cite{davis1989tam}, measuring perceived usefulness and behavioral intention as core TAM constructs, plus trust as an additional adoption factor. These items capture anticipated perceptions of a hypothetical CI recommendation system, not actual post-adoption acceptance, recognizing that these may differ. We will derive design implications for CI recommendation systems, document threats to validity (non-response bias, construct validity, limited generalizability), and assess whether results support distinct recommendation paths for CI versus non-CI users.

\smallskip
\noindent\textbf{Phase 8: Manuscript Preparation.}
We will draft Results section synthesizing quantitative and qualitative findings with figures and tables. Draft Discussion section connecting findings to prior literature (Related Work), interpreting implications, and addressing validity threats. Finalize Methods, Abstract, and Introduction.

\subsection{Ethical Considerations}
\vspace{-2pt}
This study involves human participants and will adhere to the Tri-Council Policy Statement: Ethical Conduct for Research Involving Humans~\cite{tcps2018}. All ethical safeguards detailed below will be followed to protect participant welfare, privacy, and autonomy.

\smallskip
\noindent\textbf{Research Ethics Board (REB) Approval.}
An ethics application for this study has been approved by Trent University.

\smallskip
\noindent\textbf{Informed Consent.}
Participants will provide informed consent via a Qualtrics consent page before starting the survey, covering study purpose, procedures, voluntary participation, withdrawal rights, risks, benefits, confidentiality, and contact information. Partial responses under 80\% completion will be excluded.

\smallskip
\noindent\textbf{Privacy and Confidentiality.}
We will store recruitment emails separately in encrypted, access-restricted files. Data will be transmitted via HTTPS and stored on secure institutional servers. Survey responses will be anonymous. Only aggregated, de-identified insights may be shared publicly. Data retention will follow institutional guidelines, and identifiable contact information will be deleted after study completion.

\smallskip
\noindent\textbf{Risk and Benefit Assessment.}
The survey poses minimal risk, asking only about professional experience and CI usage, with no sensitive, intrusive questions. Potential benefits include advancing understanding of CI adoption and improving CI tools. Foreseeable risks will be negligible compared to everyday professional activities.

\smallskip
\noindent\textbf{Recruitment of Participants.}
We will recruit participants via publicly available GitHub emails collected using the GitHub API and filtered for active developers. Emails will clearly state the study purpose and emphasize voluntary participation and withdrawal rights.

\smallskip
\noindent\textbf{Transparency and Open Science.}
We will share anonymized datasets, survey instrument, and analysis code after study completion, pending ethics approval, to ensure that no participants can be identified, balancing transparency with confidentiality.

\subsection{Threats to Validity}
\label{sec:validity}
\vspace{-2pt}
\noindent\textbf{External validity.}
Our sample will include only open source GitHub projects and developers with public email addresses, which may limit generalization to enterprise settings or other platforms. As in prior adoption studies, our respondents' experiences may not represent all developers~\cite{widder2019painpoints}. We will mitigate this threat by sampling diverse GitHub projects and reporting participant and project characteristics. 
We should note that GitHub will serve only as a sampling frame, not as a proxy for CI usage. CI adoption or non-adoption will be considered through participants' self-reports, since developers sampled from non-CI GitHub repositories may still report CI usage based on experience in other projects or platforms.

\smallskip
\noindent\textbf{Construct validity.}
Our findings are constrained by the survey questions we asked, which may not fully capture all aspects of CI adoption and developer perceptions. While the instrument will be piloted with 10-15 developers to identify ambiguity and survey timing, following best practices from prior surveys, some nuances may be overlooked or misunderstood, as indicated by Pinto et al.~\cite{pinto2018practices}.

\smallskip
\noindent\textbf{Internal validity.}
Self-selection bias is possible: respondents may differ from non-respondents, skewing results toward more interested or experienced developers~\cite{hilton2016usage}. As in Hilton et al.~\cite{hilton2016usage} and Widder et al.~\cite{widder2019painpoints}, recall and social desirability biases may also affect self-reported answers. We will mitigate these biases by sending reminders, offering small incentives, and monitoring demographics to adjust outreach if some groups are underrepresented.
Also, selecting developers with recent commits may favor active over occasional contributors, thus biasing our sampling.
Under-response from non-CI users may also skew the sample toward CI users, which will be monitored and addressed through targeted outreach.

\smallskip
\noindent\textbf{Conclusion validity.}
Non-response and skipped questions limit the scope of inference, so statistical associations should be interpreted cautiously; like most CI survey research, our study is best suited to highlight patterns and generate hypotheses rather than establish causal effects. We will mitigate these limitations by encouraging complete responses and interpreting results cautiously.

\section{Conclusion}
\label{sec:conclusion}
\vspace{-2pt}
This paper addresses a key gap in CI research: little is known about whether developers perceive value in CI recommendation systems or view CI as universally necessary versus context-dependent. By aiming to survey a target of 250–500 GitHub developers (CI and non-CI users alike), we will provide the first empirical evidence on developers' perceptions of such systems, including perceived value, trust, and desired features.
We make three anticipated contributions. First, we identify which adoption factors developers prioritize, distinguishing need-driven from socially-influenced motivations, service selection criteria, and switching patterns. Second, we determine whether developers view CI as universally necessary or context-dependent, revealing opinions on suitability across project types, reasoning for selective adoption, and barriers among those who view CI as context-dependent, thus informing whether recommendation systems should advise CI universally or first assess project suitability. Third, we provide empirical evidence on developers' perceptions of CI recommendation systems, testing whether the majority perceive value and whether non-CI users report higher perceived value than current users.
Our findings will benefit both researchers developing CI recommendation systems and practitioners streamlining adoption, ensuring future efforts address actual developer needs rather than assumed ones.

\bibliography{paper}
\end{document}